\documentclass{emulateapj}
 
\slugcomment{Submitted to ...}

\shorttitle{A Method for Measuring IMF Variations}
\shortauthors{Calzetti et al.}

\begin{document}

\title{A Method for Measuring Variations in the Stellar Initial Mass Function.\altaffilmark{1}}

\author{D. Calzetti\altaffilmark{2}, R. Chandar\altaffilmark{3}, J.C. Lee\altaffilmark{4}, B.G. Elmegreen\altaffilmark{5}, R.C. Kennicutt\altaffilmark{6}, B. Whitmore\altaffilmark{7}}

\altaffiltext{1}{Based on archival data (GO--10452 and GO--10501) of observations obtained with the Hubble Space
Telescope, which is operated by the Association of Universities for Research in Astronomy, Inc., 
under NASA contract NAS 5-26555.} 
\altaffiltext{2}{Dept. of Astronomy, University of Massachusetts, Amherst, MA 01003; calzetti@astro.umass.edu}
\altaffiltext{3}{Dept. of Physics and  Astronomy, University of Toledo, Toledo, Ohio}
\altaffiltext{4}{Carnegie Observatories of Washington, Pasadena, California}
\altaffiltext{4}{IBM T.J. Watson Research Center, Yorktown Heights, New York}
\altaffiltext{6}{Institute of Astronomy, Cambridge University, Cambridge, U.K.}
\altaffiltext{7}{Space Telescope Science Institute, Baltimore, Maryland}

\begin{abstract}
We present a method for investigating variations in the upper end of
the stellar Initial Mass Function (IMF) by probing the production rate
of ionizing photons in unresolved, compact star clusters with ages
$\leq 10^7$~yr and with different masses. 
We test this method by performing a pilot study on the young cluster population in the nearby galaxy NGC5194 (M51a), 
for which  multi--wavelength observations from the Hubble Space Telescope are available. Our results indicate that 
the proposed method can probe the upper end of the IMF in galaxies located out to at least $\sim$10~ Mpc, 
i.e., a factor $\approx$200 further away than possible by counting 
individual stars in young compact clusters. Our results for NGC5194 show no obvious dependence of the 
upper mass end of the IMF on the mass of the star cluster down to $\approx$10$^3$~M$_{\odot}$, although 
more extensive analyses involving lower mass clusters and other galaxies are needed to confirm this conclusion. 
\end{abstract}


\section{Introduction}

The distribution of stellar masses at birth \citep[the stellar Initial Mass Function, 
or IMF,][]{Salpeter1955} is the fundamental property, together with
the efficiency of star formation, that quantifies the conversion of
gas into stars in galaxies. It enters, directly or indirectly, in the determination of many physical
quantities of stellar populations and galaxies, such as the conversion of luminosity to 
mass and star formation rate (SFR). The IMF is often parametrized by the broken power law 
form of \citet{Kroupa2002} or the log--normal form of  \citet{Chabrier2003}.

Although many studies indicate that the stellar IMF is consistent with a universal 
form \citep[see, for a review,][]{Bastian2010},  
a few recent investigations cast  doubts on the universality in {\em all environments}. 

By comparing galaxy--integrated properties of large samples of nearby galaxies at multiple wavelengths, 
\citet{Hoversten2008}, \citet{Meurer2009}, \citet{Lee2009}, and \citet{Boselli2009} find that, on average,  
the ionizing/non--ionizing flux ratio of  galaxies decreases for decreasing galaxy luminosity or mass or SFR or SFR/area. 
\citet{Lee2009} exclude, based on statistical arguments, that galaxy--integrated stochastic sampling or variations in the star formation histories alone can fully account for the observed trend. A  systematic variation of the high end of the IMF with some galactic or local parameter (luminosity, SFR, SFR/area, gas density) has been suggested as one of the possible ways to explain the observations \citep{Krumholz2008,Meurer2009,PflammAltenburg2009,Lee2009}, although other scenarios are possible \citep[e.g.,][]{Dong2008,Kotulla2008,Gogarten2009,Hunter2010,Goddard2010}.  

The model suggested by \citet{PflammAltenburg2009} invokes   
the existence of two correlations: one between a galaxy's SFR and its clusters' maximum mass  \citep[e.g.,][]{Larsen2002}, and another (deterministic) between a cluster mass and the most massive {\em possible} star the cluster can form \citep{Weidner2006}. The nature of the second correlation  implies 
that clusters are not subject to the effect of stochastic sampling of the IMF \citep{Elmegreen2006}, i.e. 100 clusters each with mass=10$^2$~M$_{\odot}$ would not be equivalent to one 10$^4$~M$_{\odot}$ cluster, as the former will systematically lack the  high--mass stars contained in the latter.

The standard approach to deriving  the functional form of an IMF, i.e. counting the individual stars contained in single--age, young stellar populations (typically  $\lesssim$3~Myr~old star clusters), 
is difficult to apply over statistically large samples of clusters at a given mass and in a variety of galactic environments, and 
provides limited leverage for discriminating stochastic sampling from systematic trends in the IMF. 
Significant blending will occur for massive stars due 
to crowding in the centers of star clusters \citep{Apellaniz2008,Ascenso2009}, even with 
the angular resolution of the Hubble Space Telescope, for distances larger than those of the Magellanic 
Clouds ($\sim$55~kpc).  These difficulties may account for the contrasting results obtained by \citet{Maschberger2008}, who concluded that the most massive stars in Milky-Way clusters
are stochastically drawn from a universal IMF, and \citet{Weidner2010}, who claimed evidence
for a systematic trend between the masses of clusters and those of their most massive stars. 

Here we present an alternate method that exploits the integral properties of star clusters to place limits on 
variations of the high--end IMF. Each star cluster is treated as single unit, and IMF variations 
are investigated by measuring the dependence of the  ionizing photon rate, normalized by the cluster mass, 
on cluster mass (section~2). This method 
extends the \citet{Corbelli2009}  approach (the cluster birthline) by normalizing the ionizing photon rate to 
the age--independent cluster mass, instead of the age--dependent bolometric luminosity. Our method, 
however, requires that accurate ages for each cluster be determined. 
We test our approach using observations of the young star cluster population in the nearby galaxy NGC5194 
(section~3). A summary of findings and future directions are given in section~4. 

\section{Method and Predictions}

Stellar population synthesis models provide specific predictions for the ionizing photon rate, N(H$^o$), produced by a 
single--age stellar 
population as a function of time, for a given IMF, mass, and metallicity. Compact star clusters represent a 
close approximation to the idealized case of a  single--age population. The rate  N(H$^o$) decreases with 
increasing age, as massive stars die out, and ionized gas generally becomes undetectable 
in clusters older than $\sim$8--10~Myr (more than 100 times fainter than at 1--2~Myr age). Most HII regions 
form expanding   
shells around the central cluster \citep{Martin1998}, thus decreasing the surface brightness of the ionized gas emission lines even further. For practical purposes, ionized gas emission will not be easily detected from star clusters 
older than $\approx$6--8~Myr. 

Measuring N(H$^o$) from a star cluster at a given (young) age is equivalent to measuring 
the number of massive stars distributed according to an IMF. This, in models, includes fractional stars in the high end of 
the IMF, for low--mass clusters. Real clusters, however, 
do not contain fractions of stars, and, for masses$\lesssim$a~few$\times$10$^4$~M$_{\odot}$, they will be subject to 
stochastic sampling of 
the IMF, even if the underlying IMF is `universal', i.e., independent of cluster mass \citep{Elmegreen2006}. Stochastic 
sampling induces a dispersion on the number of stars at a given stellar mass that are present in a real cluster, and the dispersion increases for decreasing cluster mass. This, in turn, affects a number of observational parameters, 
including  the inferred mass and N(H$^o$) from clusters \citep[e.g.,][]{Cervino2002,Cervino2004}. The scatter on the mass determination increases from  6\% for a 10$^5$~M$_{\odot}$ cluster to $\sim$50\% for a 
10$^3$~M$_{\odot}$ cluster; the scatter on  N(H$^o$) increases from $\sim$10\% to 80\% for the 
same cluster mass range \citep{Cervino2002}. In our approach, the impact of stochastic IMF sampling is minimized by combining large numbers of same--mass, young clusters to a cumulative equivalent mass $\approx$10$^5$~M$_{\odot}$, since a `universal' IMF would be fully sampled in a cluster of this mass, and stochastic uncertainties are reduced 
to 10\% or less in both M$_{cl}$ and  N(H$^o$).

Figure~\ref{fig1}, left,  shows the model expectations for N(H$^o$) (as traced by the H$\alpha$ luminosity, L(H$\alpha$)) per unit cluster mass, as a function of cluster mass, for both cases of a `universal' IMF \citep{Kroupa2002} with fractional stars and a cluster--mass--dependent IMF with the maximum possible stellar mass depending on the cluster mass \citep[][]{Weidner2006}. Results similar to those of \citet{Weidner2006} would be obtained if the IMF became steeper at the high end 
for decreasing cluster mass. 
The L(H$\alpha$)/M$_{cl}$ ratio is averaged over the age ranges 1.5--5~Myr and 2--8~Myr for constant star formation, in order to bracket the uncertainty in cluster ages found in most observational analyses. 
While our default model has metallicity 1.5~solar (to match NGC5194, see below), we show examples also at 1/3 and 3.5 solar metallicity. The metal content of the ionized gas has, in fact,  strong influence on the ratio L(H$\alpha$)/M$_{cl}$;  this parameter will be critical when comparing star cluster populations belonging to different galaxies, while its impact will be less important when comparing clusters at different masses {\em within} the same galaxy, under the reasonable assumption that no systematic trend exists between a cluster 
mass and its metal content. 

Stochastic sampling of the stellar IMF in low--mass clusters, combined with photometric uncertainties and reddening, can also affect age determinations from multi--wavelength data. Clusters younger than $\sim$10~Myr that do not have ionized gas emission can be assigned older ages; conversely, clusters older than 10~Myr can be assigned ages $\sim$6--9~Myr 
\citep{Chandar2010b,Fouesneau2010}.
Thus samples of young, low--mass clusters are  `polluted' by either addition of old clusters or loss of young members. The two  may compensate each other to some degree, but  this is still unquantified. Furthermore, masses may be underestimated by up to a factor of 2 \citep{Fouesneau2010}. 
We discuss both effects below,  when applying our approach to actual data. 

\section{Application: a Test Case}

We test our method on the nearby star--forming 
galaxy NGC5194 \citep[D$\sim$8.4~Mpc,][]{Tully1988}, for which extensive multi--wavelength, broad-- and narrow--band,  archival imaging from the Hubble Space Telescope exists (GO--10452 and GO--10501). The galaxy represents an ideal 
target for this test: it is nearly face--on ($i\sim$20$^o$), thus minimizing line--of--sight overlaps among clusters, and its cluster population has been extensively studied \citep[e.g.][]{Bastian2005,Chandar2010b}. The HST observations of NGC5194 include U (with WFPC2), BVI and H$\alpha$ (with ACS) for the entire disk, and P$\alpha$($\lambda$=1.8756~$\mu$m, with NICMOS) for the inner $\sim$3~kpc radius \citep{Scoville2001}. The galaxy has 
a mean metal abundance about  twice solar, which remains super--solar across the entire disk up to R$_{25}$ \citep{Moustakas2010}.

We use the sample of clusters identified by \citet{Chandar2010b}, and refer the reader to that paper for the sample selection and measurements. Descriptions of the selection and age--dating methods for star cluster populations in external galaxies  can be found also in \citet[][applied to the Antennae Galaxies]{Whitmore2010} and \citet[][applied to NGC5236]{Chandar2010a}. We summarize here some of the steps that are relevant to the present paper. For all clusters, 
ages and extinctions are determined from the $\chi^2$ minimization between the observed U,B,V,I and H$\alpha$ (non--continuum--subtracted) photometry and predictions from the \citet[][2007 Update]{Bruzual2003} stellar population models combined with a Milky--Way--type extinction curve \citep{Fitzpatrick1999}. The choice of the extinction curve does not impact the results, because most curves are degenerate in the wavelength range from U to I \citep{Calzetti2001}. Changing 
the population synthesis models, e.g., to Starburst99 \citep{Leitherer1999}, also has minimal impact on the results in the case of young clusters ($\lesssim$10~Myr). The use of the non--continuum--subtracted H$\alpha$ photometric point (which becomes a R--band point in the absence of ionized gas) adds leverage for separating young, extincted clusters from old, relatively unextincted ones, even in those cases where the young clusters do not ionize gas 
\citep{Chandar2010b}. 
The typical uncertainty is a factor $\sim$2 in the final age determination, and  $\sim$60\%  in the mass. 
The mass of each cluster is estimated from its
extinction-corrected $V$ band luminosity and the age-dependent mass-to-light ratio 
from the stellar population models, assuming a Kroupa/Chabrier stellar IMF. Changing the upper mass limit of the IMF from 120 to 30~M$_{\odot}$ changes the mass estimate by $\approx$2.5X, which is within the range of uncertainties we consider. We consider  two mass bins, located at the two extremes of the sampled range in the cluster mass distribution: $\sim$1.5$\times$10$^3$~M$_{\odot}$ and $\sim$4$\times$10$^4$~M$_{\odot}$. They span about a factor 30 in mean mass and a factor 300 in individual cluster masses.

We inspect each cluster visually, and remove from the sample any cluster located in a sufficiently crowded region to hamper H$\alpha$ line measurements. Our final, visually inspected, sample consists of  86 clusters in the mass range 
0.5--2~10$^3$~M$_{\odot}$ and 57 in the range 2--15~10$^4$~M$_{\odot}$, all younger than 10~Myr. Of the 86 clusters 
in the lowest mass bin,  39 (45\%) are undetected in H$\alpha$ down to our 3~$\sigma$ limit of 7$\times$10$^{36}$~erg~s$^{-1}$ (calculated in a 5 pixels radius aperture~=~10~pc). As mentioned in section~2, some young clusters can be assigned 
ages older than 10~Myr, and vice--versa. 
We address this issue by adopting two extreme scenarios: case~(a): all clusters whose colors indicate young ages  are included in the average of each mass bin; case~(b):  only clusters which have colors of young populations {\em and}  also have H$\alpha$ detections above 3~$\sigma$ are included in the mass bin average. In order to account for the possibility that cluster masses in the low--mass bin may be underestimated by as much 
as a factor 2 \citep{Fouesneau2010}, we include clusters with masses as low as 500~M$_{\odot}$.  Of the 57 clusters in the higher mass range, only 6 (11\%) are below 3~$\sigma$ in H$\alpha$. 

Photometric measurements of the hydrogen recombination lines (continuum--subtracted H$\alpha+$[NII] and P$\alpha$) are performed at the positions of the clusters,  
with aperture sizes that scale according to the cluster's mass, i.e., according to the expected Str\"omgren radius of the HII region. Aperture sizes are chosen to be about 0.35~R$_{Stroemgren}$, because of crowding of the clusters, and age--dependent aperture corrections are implemented based on the few `isolated' HII regions identified in our sample. The final photometry is visually inspected to check for contamination from neighboring sources; the local background, determined for each aperture from annuli 2~pixels wide around each region, is subtracted from each photometric measurement, in order to avoid contamination by diffuse emission not associated with the cluster. The uncertainty in the final H$\alpha+$[NII] flux from uncertainties in the aperture correction range from about 20\% for the youngest and most luminous regions to  about 80\% in clusters older than $\sim$5--6~Myr.  
[NII] contamination in the narrow--band filter  
is removed by using the average galactic [NII]/H$\alpha$ ratio \citep{Moustakas2010}.  The P$\alpha$ image only covers a portion of the NGC5194 disk, and is used here to derive ionized gas extinction corrections region by region for the area of overlap between the ACS and NICMOS mosaics \citep[see, also,][]{Scoville2001}, for a total of 29  regions. The median values of the gas extinction corrections in each cluster mass bin are then adopted for the regions outside of the overlap area. This may lead to an overestimate of the dust extinction correction of H$\alpha$ in each mass bin, as the mean extinction shows a gradient of decreasing values towards the outer regions \citep{Calzetti2005}; this will not impact the relative comparison between mass bins, only their absolute values.

\begin{figure}
\figurenum{1}
\plottwo{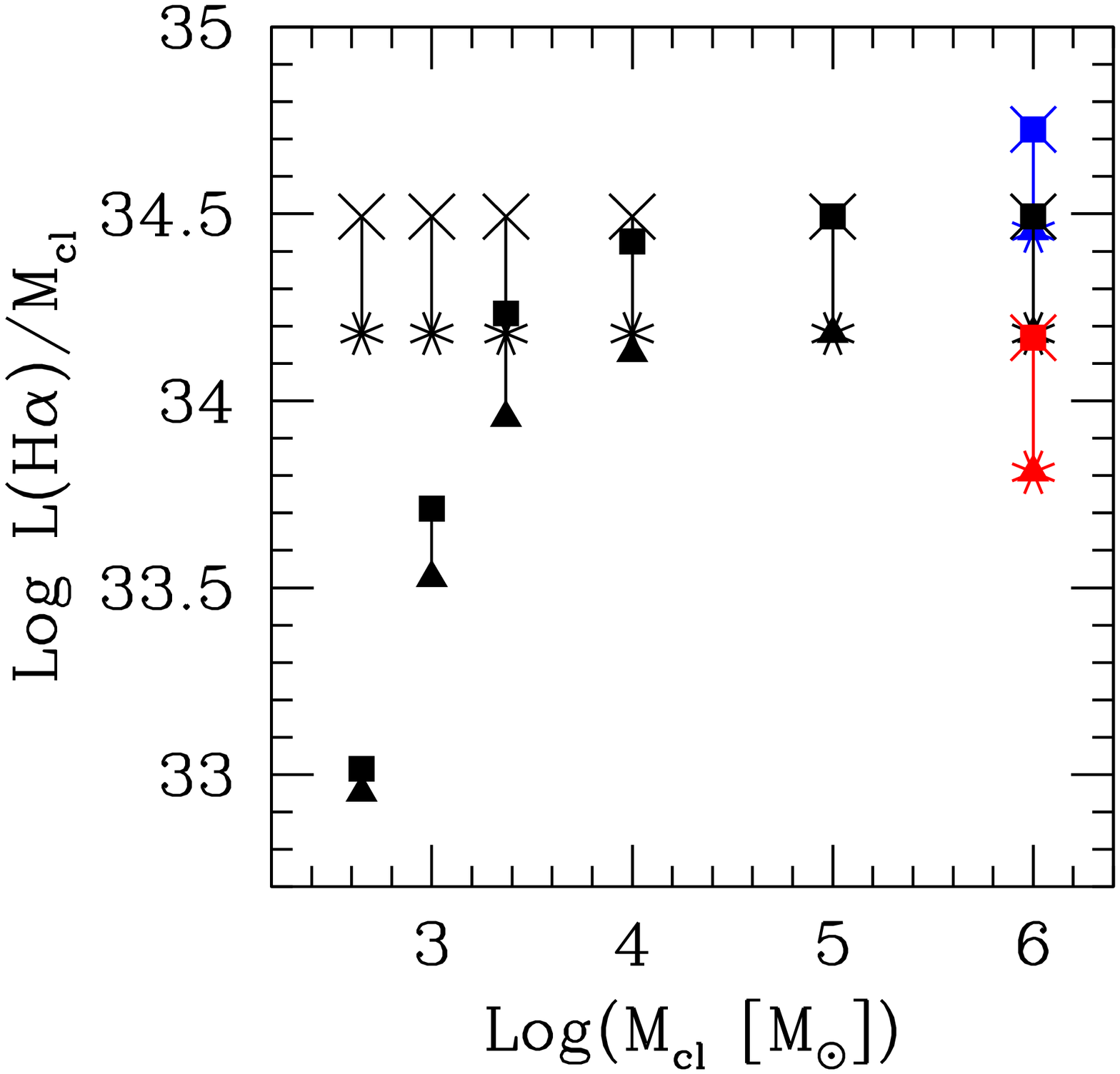}{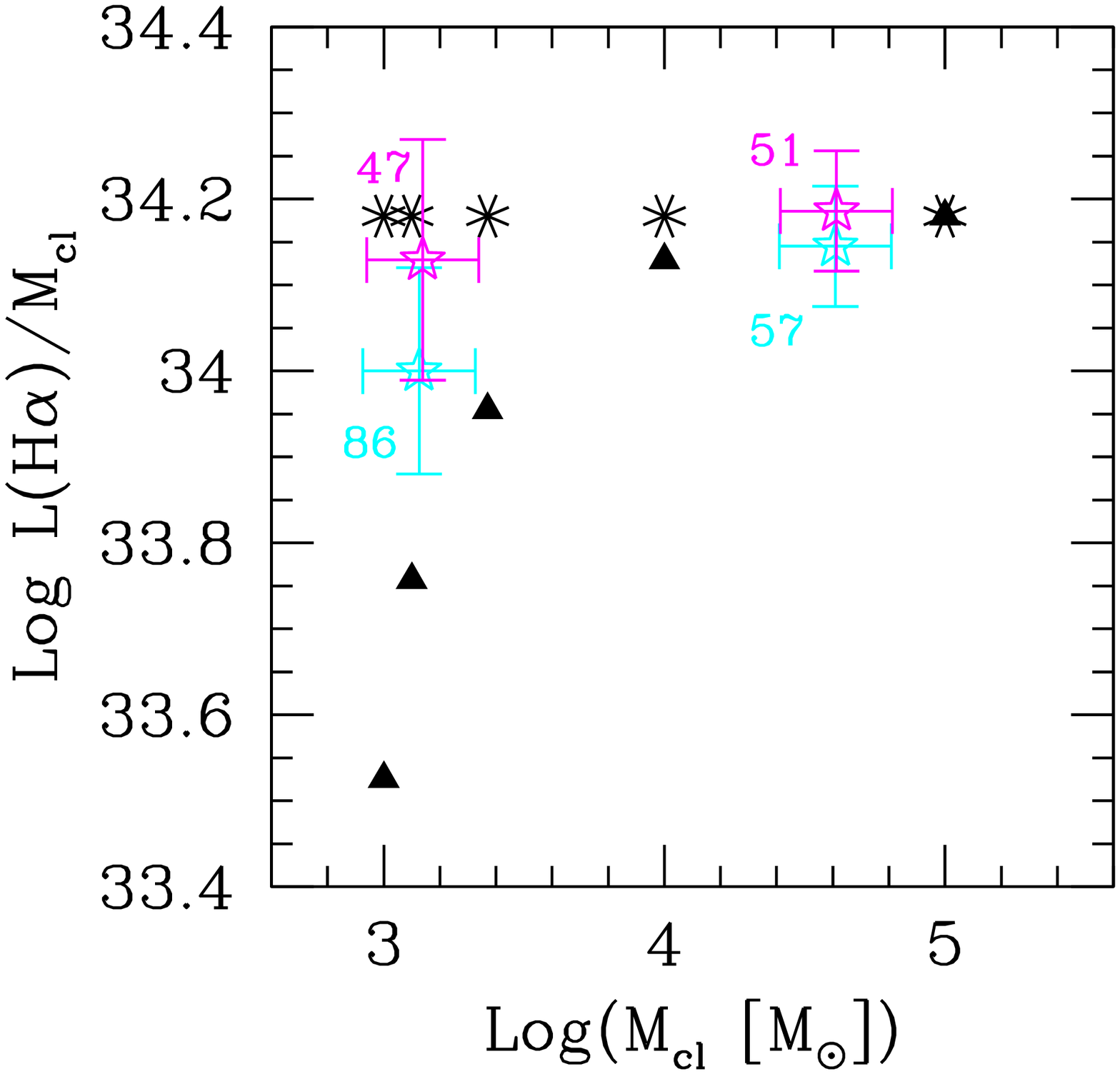}
\caption{{\bf LEFT:} The expected ratio L(H$\alpha$)/M$_{cl}$ as a function of the cluster mass M$_{cl}$ both 
for a universal IMF (crosses and asterisks), and for a cluster--mass--dependent IMF \citep[filled squares and triangles,][]{Weidner2006}. The models are from Starburst99 \citep[2007 Update,][]{Leitherer1999}, and assume averages over large numbers of clusters. The mean L(H$\alpha$)/M$_{cl}$ ratios are averaged over two age ranges: 1.5--5~Myr (crosses and squares) and 2--8~Myr (asterisks and triangles), joined by vertical bars, under the assumption of constant star formation for each age range. We assume that clusters younger than 1.5--2~Myr are too dust--enshrouded to provide significant contribution at optical wavelengths. The black symbols are for 1.5~solar metallicity models; examples of higher ($\sim$3.5 solar, red) and lower ($\sim$1/3 solar, blue) metallicity are shown at a representative mass. 
 {\bf RIGHT:} The ratio $<$L(H$\alpha$)$>$/$<$M$_{cl}>$ as a function of the mean cluster mass M$_{cl}$ in two mass bins
  for NGC5194, for clusters younger than $\sim$8--10~Myr (stars with 1~$\sigma$ error bars), including all clusters within each mass bin (cyan, bottom symbols), and including only clusters with H$\alpha$ flux detections $>$3~$\sigma$ (magenta, top symbols). The number of clusters in each  mass  bin is shown close to the symbols.  Expectations from models with 1.5 solar metallicity and averaged over the 2--8~Myr age range are shown (asterisks and triangles), as they are the closest match to the NGC5194 high--mass data. 
\label{fig1}}
\end{figure}

In Figure~\ref{fig1}, right, the ratio L(H$\alpha$)/M$_{cl}$=$<$ L(H$\alpha$)$>$/$<$M$_{cl}$$>$ for each of the two cluster mass bins is shown for the two color--age cases (a) and (b), as cyan and magenta points, respectively. 
The error bars of the high--mass bin are dominated by photometric and aperture--correction uncertainties, while those of 
the low--mass bin contain a significant contribution from the stochastic uncertainty in cluster mass determinations. 
The absolute values of  L(H$\alpha$)/M$_{cl}$ are similar for the two mass bins, within the 1~$\sigma$ uncertainty, for both sets of points. The absence of an obvious differential trend in L(H$\alpha$)/M$_{cl}$ for the two mass bins is consistent with a universal, stochastically--sampled,  IMF. 
However, we cannot completely eliminate the scenario for which the maximum possible mass of a star increases with the cluster mass, as the data are only 1--2~$\sigma$ away from this model's expectations (triangles). 

The data follow the 2--8~Myr age average points (asterisks) of 
the 1.5~solar~metallicity stellar population models, implying the following, non--exclusive, possibilities: (1) our criteria select clusters fairly uniformly distributed in the age range 2--8~Myr; (2) the models used in Figure~\ref{fig1}, right, have too low metallicity for NGC5194, and a higher metallicity model should be used (cf. Figure~\ref{fig1}, left panel); (3) significant, mass--independent, ionizing photon leakage is present in our regions, which we have not attempted to correct for in our analysis. 

\section{Conclusions and Future Work}

We have presented a method to test for variations of the high--end of the stellar IMF that exploits the integrated properties of star clusters, rather than resolved star counts. This approach can overcome statistical limitations in IMF determinations and average out effects of stochastic sampling by leveraging the large numbers of star clusters in galaxies. The ratio of the ionizing photon rate (as measured by the extinction--corrected H$\alpha$ emission) from  young clusters to cluster mass  provides a powerful tool to test for such variations.  A basic requirement for this method is to isolate the young ($<$10~Myr) star cluster population, and measure its ionized gas emission. Accurate ($\sim$2X) ages can be obtained with high--angular--resolution (e.g., HST) multi--color observations; a minimum set appears to be UBVRI \citep{Chandar2010b}, although the addition of the UV can add an extinction leverage \citep{Calzetti2005} to further reduce the age uncertainty. Ionized gas emission measurements also need to be corrected for dust extinction, requiring observations of two hydrogen recombination lines. Comparisons between galaxies further requires that the metallicity of each galaxy is known, as the ionizing photon rate is 
highly dependent on metallicity. 

We have tested this method by performing a pilot study on the cluster population in the nearby galaxy NGC5194. 
Our analysis shows that, when a sufficiently large number of similar--mass clusters is 
summed together to average out effects of stochastic sampling of the IMF, the mean H$\alpha$ luminosity normalized 
by the mean cluster mass is relatively independent of the cluster mass, down to M$_{cl}\sim$10$^3$~M$_{\odot}$. This is  consistent with an effectively universal IMF independent of the cluster mass. We do not find strong evidence that the
 maximum possible mass of a star in a cluster decreases systematically with decreasing cluster 
 mass \citep[as proposed by][]{Weidner2006}. 
However, this is not the most stringent test of the stochastic sampling model because clusters more massive than 10$^3$~M$_{\odot}$ are expected to contain more than 1 O--type star.  
Clearly, statistics need to be expanded by including lower cluster masses, and cluster populations from other galaxies, spanning a range of internal and external environments. The proposed method enables tests of variations of the high--end IMF up to at least $\sim$10~Mpc distance (NGC5194 is located at 8.4~Mpc distance), and possibly more, depending on the crowding of the cluster population. This distance range is at least a factor of 200 larger than what has been achieved so far with direct star counts. 

\acknowledgments

This research has made use of the NASA/IPAC Extragalactic Database (NED) which is operated by the Jet Propulsion Laboratory, California Institute of Technology, under contract with the National Aeronautics and Space Administration.

\end{document}